\documentclass[article,12pt]{article}
\usepackage[utf8]{inputenc}
\usepackage[T1]{fontenc}
\usepackage[colorinlistoftodos]{todonotes}
\usepackage[letterpaper, margin=1in]{geometry}
\usepackage{authblk}
\usepackage{mathptmx}
\usepackage{amsmath,amssymb,amstext}
\usepackage{setspace}
\usepackage[margin=1in]{geometry}
\usepackage{enumitem}
\usepackage[square,sort,comma,numbers]{natbib}

\usepackage{chngcntr}
\counterwithout{subsubsection}{subsection}

\newcommand{\la}{\raise0.3ex\hbox{$<$}\kern-0.75em{\lower0.65ex\hbox{$\sim$}}}
\newcommand{\ga}{\raise0.3ex\hbox{$>$}\kern-0.75em{\lower0.65ex\hbox{$\sim$}}}

\renewcommand\refname{{}}

\date{\vspace{-5ex}} 

\title{{\sc -- Plasma 2020 --} \\
{\large\sc White Paper for the Decadal Assessment of Plasma Science}\\
---\\
{\bf Intracluster Medium Plasmas}}

\author[1]{Damiano Caprioli\thanks{E-mail: caprioli@uchicago.edu; Phone: +1 (773) 834-0824}}
\affil[1]{\small University of Chicago, Department of Astronomy \& Astrophysics, Chicago, IL 60637, USA}

\author[2]{Gianfranco Brunetti}
\affil[2]{\small INAF, 
Istituto di Radioastronomia,  Bologna, BO 40129, Italy}

\author[3]{Thomas W. Jones}
\affil[3]{\small University of Minnesota, School of Physics and Astronomy, Minneapolis, MN, 55455, USA}

\author[4]{Hyesung Kang}
\affil[4]{\small Pusan National University, Department of Earth Sciences, Busan, 46241, Korea}

\author[5]{Matthew Kunz}
\affil[5]{\small Princeton University, Department of Astrophysical Sciences, Princeton, NJ, 08544, USA}

\author[6]{S. Peng Oh}
\affil[6]{\small University of California, Santa Barbara, Department of Physics, Santa Barbara, CA 93106, USA}

\author[7]{Dongsu Ryu}
\affil[7]{\small UNIST, Department of Physics, Ulsan, 44919, Korea}

\author[1]{Irina Zhuravleva}

\author[9]{Ellen Zweibel}
\affil[9]{\small  University of Wisconsin--Madison, Department of Astronomy, Madison, WI 53706, USA}

\affil[*]{Primary author}

\begin{document}

\maketitle

\newpage
\pagenumbering{arabic}

\subsubsection{The Fundamental Role of Plasma Physics in Galaxy Clusters}
\vspace{-3mm}
Galaxy clusters are the largest and most massive bound objects resulting from cosmic hierarchical structure formation. They span several Mpc (${\sim}10^{20}~{\rm km}$) with total masses ${\lesssim}10^{15}~{\rm M}_{\odot}$ (${\lesssim}10^{45}~{\rm kg}$). Baryons account for somewhat more than 10\% of that mass, with roughly 90\% of the baryonic matter distributed throughout the clusters as hot ($T>1$ keV), high-$\beta~(\ga 10-10^3$), very weakly collisional ($\lambda_{\rm Coulomb}\gtrsim 10^{14}~\lambda_{\rm Debye} \sim 10^{11}~c/\omega_{\rm pi}$) plasma; the so-called ``intracluster medium'' (ICM). 
Cluster mergers, ``close gravitational encounters'' and accretion, along with violent feedback from galaxies and relativistic jets from Active Galactic Nuclei (AGNs), drive winds, gravity waves, turbulence and shocks within the ICM. 
Those dynamics, in turn, generate cluster-scale magnetic fields and accelerate and mediate the transport of high-energy charged particles (\emph{cosmic rays}, CRs). 
Kinetic-scale, collective plasma processes define the basic character and fundamental signatures of these ICM phenomena, which are observed primarily by X-ray and radio astronomers. 

The ICM thermal emission typically falls in the X-ray band. By measuring its spectrum and intensity both plasma temperature and density can be inferred; 
furthermore, X-ray lines and high-resolution spectroscopy can provide information about turbulent motions and deviations from local thermodynamic equilibrium.
The ICM can be regarded as a laboratory to test theories of heat transport, as well as viscous and resistive dissipation, and to measure turbulence properties over a broad range of physical scales that encompass nominal particle-particle interaction lengths. 

In recent years broad theoretical efforts, both analytic and simulation (hydro, MHD, and kinetic), have helped to unravel the central roles of plasma processes in shaping the ICM dynamics, and especially in determining kinetic-scale properties that control astrophysically-important and potentially diagnostic cluster-scale behavior such as heat transport, magnetic-field generation, and CR acceleration. The ICM is a  
%
%
promising venue for the study of fundamental processes in collisionless plasmas such as shocks and magnetic turbulence. 
Astrophysicists connect these phenomena to the generation of energetic particles and hence to \emph{non-thermal} emission that spans from radio, to X-rays, and possibly even to $\gamma$-rays, as a consequence of processes that are both leptonic (synchrotron, bremsstrahlung, and inverse-Compton scattering) and hadronic (decay of $\pi_0$ produced in nuclear interactions involving relativistic protons).

Modern telescopes are revealing unprecedented information about the complexity of  high-$\beta$ plasmas, and the future generation promises to open additional new windows to probe it.
At the same time, theoretical studies and observations raise intriguing questions about actual ICM plasma conditions and challenge the current paradigm for ICM dynamics and its observable signatures. 

\begin{figure}
    \centering
    \includegraphics[width=0.8\textwidth, clip=true, trim= 0 0 0 11]{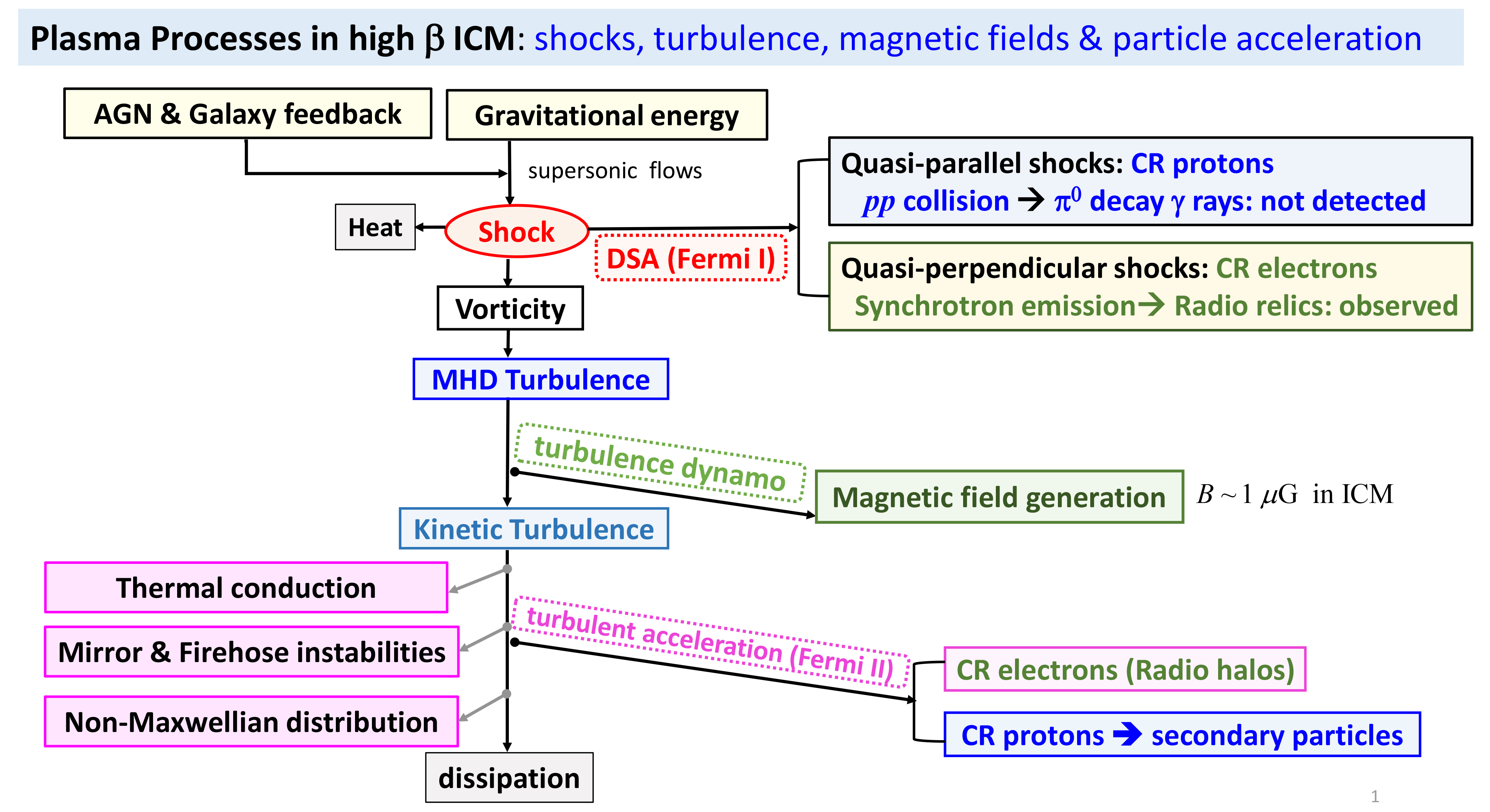}
        \vspace{-4mm}
    \caption{\small Key physical processes governing high-$\beta$ ICM plasma and their observational signatures.}
        \vspace{-5mm}
    \label{fig:ICM_physics}
\end{figure}

In this white paper, we argue that plasma physics is vital to understanding galaxy clusters and that the exceptional properties of the ICM provide a unique opportunity to expand our understanding of fundamental plasma physics and its role in the evolution of the Universe (see Fig.~\ref{fig:ICM_physics}).

\begin{figure}
    \centering
    \includegraphics[width=1\textwidth]{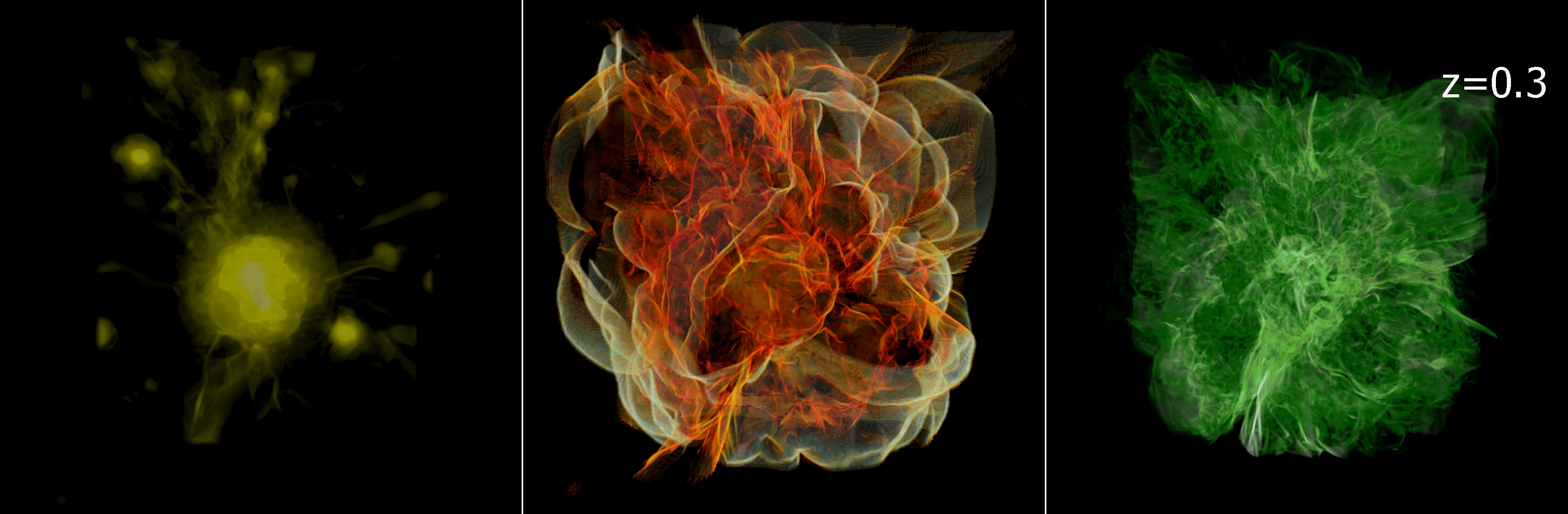}
    \vspace{-9mm}
    \caption{\small Left to right: Simulated ICM density, shocks (red $M \lesssim 3$, yellow $M\sim 10$), and flow vorticity, a proxy for turbulent intensity. Shocks and turbulence are ubiquitous, especially during cluster mergers \cite{vazza2017}.}
    \label{fig:sims}
    \vspace{-5mm}
\end{figure}

\vspace{-5mm}
\paragraph{Shocks in the Intracluster Medium.}
Hierarchical clustering of the large-scale structure of the Universe induces supersonic flows, leading to the formation of weak shocks in the ICM \cite{Ryu2003,vazza2017} (Fig.~\ref{fig:sims}).
Shocks are expected to play key roles in governing the plasma physics of the ICM: turbulence generated behind shocks can amplify magnetic fields through turbulent dynamo and accelerate particles via the Fermi-II process \cite{Ryu2008,Bruggen2011}.
Similar to shocks in the heliosphere and in supernova remnants, ICM shocks are thought to accelerate CR ions and electrons via diffusive shock acceleration (DSA, Fermi-I process) \cite{caprioli-spitkovsky14a,caprioli+17,brunetti-jones14}.
Although the acceleration of CR electrons can be inferred from the so-called radio-relic shocks detected in merging clusters \cite{vanweeren2010}, the presence of CR protons produced by ICM shocks has yet to be confirmed.
Inelastic collisions of CR protons are expected to produce diffuse hadronic $\gamma$-ray emission from clusters, which, somewhat unexpectedly, has not yet been detected \cite{Ackermann2016}, marking a sharp distinction from $\gamma$-ray-bright supernova shocks.

In the past decade, kinetic simulations (both particle-in-cells, PIC, and hybrid, i.e., kinetic ions/fluid electrons) have shed new light on the complex non-linear interplay between electromagnetic fields, non-thermal particles, and thermal plasma.
Hybrid simulations have greatly refined the theory of ion DSA, unraveling how particles are injected into DSA from the thermal bath, how they drive instabilities that foster their own scattering, and how non-thermal \emph{seeds} can be re-accelerated by the shock passage \cite{caprioli-spitkovsky14a,caprioli-spitkovsky14b,caprioli+18}.
PIC simulations have also shown simultaneous DSA of ions and electrons \cite{park+15} and suggested that DSA is suppressed for low sonic Mach number, $M_s$ \cite{Ha2018}.

\vspace{-5mm}
\paragraph{Intracluster Turbulence.}
Turbulence in the ICM is driven by ongoing accretion and mergers, as well as by galactic winds and AGN feedback \cite{brunetti-jones14}. 
Simulations of cosmic structure formation have shown that such turbulence is mildly trans-sonic with $M_s \sim 0.5$ \cite{Ryu2008,vazza2017}, consistent with X-ray observations \cite{churazov2012,simionescu+19}.
ICM turbulence is important in redistributing heavy elements and entropy, gas heating, particle acceleration and transport, and likely in non-thermal ICM pressure support. 
In addition, since the ICM is highly ionized and conducting, such turbulence may amplify even very weak seed magnetic fields through the small-scale, turbulent dynamo \cite{Ryu2008, Porter2015}.

High-resolution X-ray images have enabled significant progress in our understanding of transport processes in the ICM. 
In many clusters, the measured width of the interface between contact discontinuities (``cold fronts'') and hot ICM gas is narrower than $\lambda_{\rm Coulomb}$, suggesting that diffusive processes are locally suppressed \cite{2016JPlPh..82c5301Z}. 
Very deep Chandra observations of the Coma cluster allowed to probe fluctuations in the bulk ICM plasma on $\lambda_{\rm Coulomb}$ scales and have showed that the power spectrum of fluctuations follows a Kolmogorov-like scaling, implying that the effective gas viscosity in the bulk ICM is suppressed to be at least an order of magnitude less than $\lambda_{\rm Coulomb}$-based, Spitzer viscosity [Zhuravleva et al., submitted]. 
The observed increased effective collision rate in Coma could be due to particle scattering off fluctuations produced by kinetic-scale plasma instabilities \cite{kunz+14} or it could reflect transport processes that are anisotropic (perhaps with respect to the local magnetic-field direction  \cite{braginskii65}).

Assuming a statistical linear relation between measured density fluctuation amplitude and ICM turbulent velocity, the velocity power spectra have been measured within so-called ``cool cores'' of galaxy clusters. 
Typical velocities are ${\gtrsim}100~{\rm km/s}$ on scales of ${\sim}50~{\rm kpc}$ and the slopes of spectra are consistent with Kolmogorov  \cite{2014Natur.515...85Z,2018ApJ...865...53Z}. 
These results are similar to those measured directly through line broadening and shift using the {\it Hitomi} X-ray observation of the Perseus cluster \cite{hitomi2016}. 

ICM turbulence is also very likely to be important for stochastic CR acceleration, despite being slow, so acting on cosmological time-scales. Models for stochastic CR acceleration as well as long-range CR transport still need refinement and verification, being dependent upon the microphysical plasma and macrophysical magnetic field details \cite{bb05,bl07}.
In this respect, it is worth recalling that CR protons are trapped in clusters for times comparable to the age of the universe, and relativistic electrons may survive for $0.1$--$1~{\rm Gyr}$;
therefore, the ICM is a unique probe of the slowest acceleration mechanisms in the Universe. 

\vspace{-5mm}
\paragraph{Heat Transport in the ICM.}

In the ICM it is possible to single out heat sinks (spatially resolved radiative cooling) and heat sources (e.g., AGN), but the transport of energy within the ICM remains controversial. One such process, thermal conduction, has important effects on gas thermodynamics in fusion, space and astrophysical plasmas. 
In clusters, it affects local and global thermal stability and the damping of acoustic modes. However, its potency remains highly uncertain; the widely assumed Spitzer heat flux assumes collisionality dominates. 
In the solar wind, where $\beta \sim 1$, there is rough agreement with Spitzer \cite{bale2013electron}, but the high-$\beta$, weakly collisional ICM plasma is an excellent laboratory for probing instabilities that can inhibit the field-aligned heat flux, $q_{\parallel}$, below Spitzer values.
Two such candidates actively being investigated with kinetic simulations are: 
 (i) the mirror instability, driven by pressure anisotropy, where a modest suppression ${\sim}0.2$--$0.3$ is predicted \cite{kunz+14,komarov16};
 (ii) whistler waves driven by the heat flux itself, where $q_{\parallel} \sim 1/\beta$ independent of temperature gradient  \cite{gary00,roberg-clark18,komarov18}. 
 Energy transport in the ICM can also be mediated by turbulent gas motions
 or via CRs, which can transfer energy to the gas via a gyroresonant streaming instability \cite{guo08-CR,ruszkowski17} (although the transport of CRs in the ICM is similarly under debate \cite{wiener13a}). 
Kinetic phenomena may also control thermal conduction \cite{Roberg-Clark2016},
the damping of Alfv\'en waves in high-$\beta$ plasmas \cite{squire+17} and viscous dissipation \cite{kunz11,Komarov2018}.

X-ray images of bright clusters also allow us to infer the effective equation of state of perturbations in the ICM, revealing that isobaric fluctuations are energetically dominant in the cores and adiabatic fluctuations contribute less than 10\% to the total perturbation variance  \cite{2016ApJ...818...14A,2016MNRAS.458.2902Z,2016MNRAS.463.1057C}.

\vspace{-5mm}
\subsubsection{Future Challenges and Open Questions}
\vspace{-3mm}
\begin{itemize}[noitemsep,leftmargin=*]
\item {\bf Origin of cosmic magnetic fields.} Can dynamo and AGN feedback explain their strength and scales? 
What are the nature and properties of fields in filaments of galaxies and in cosmic voids?

\item {\bf Shock acceleration.}
How sensitive is shock acceleration to the plasma $\beta$ and the sonic Mach number?
Are ICM shocks too weak (subcritical) to inject ions into DSA?
How rapidly can collisionless processes drive ions and electrons to thermal equilibrium behind a shock?
Is the lack of $\gamma$-ray emission from clusters at odds with DSA predictions? 

\item {\bf Turbulence.} How is ICM turbulence generated? What are its basic properties (amplitude, spectrum, anisotropy, injection scales)?  How do density and magnetic fluctuations relate? 
How is energy dissipation partitioned into heating, CR acceleration, and magnetic fields? 

\item {\bf Turbulent re-acceleration.} 
Can turbulent re-acceleration explain all the properties of cluster-scale radio emission? What are the sources of seed electrons? Are shocks viable candidates? Can we predict the slope and normalization of re-accelerated electrons via kinetic simulations?

\item {\bf Particle distributions.}
Do ion/electron time-stationary distributions deviate from Maxwellian? May $\kappa$ distributions provide a better fit? Are they a local or global phenomenon?

\item {\bf Energy Transport.} 
How does energy transport proceed from cluster-size scales  down to the micro-scales responsible for heating and electron re-acceleration?
What are the essential properties of viscosity and conduction in weakly-collisional, high-$\beta$ plasmas?
What is the influence of plasma micro-instabilities?
How anisotropic are such processes with respect to the local magnetic field? How can one probe these effects with current and future observatories?
\end{itemize}

\vspace{-7mm}
\subsubsection{Current and Next-generation Observatories}
\vspace{-3mm}

\hspace{6mm}{\bf Radio.}
The innovative LOFAR radio telescope provides a breakthrough in sensitivity at very low (meter) wavelengths, opening a new observational window \cite{vanharlem+13}. 
Low-frequency observations are particular crucial because they target ${>}100$-Myr-old electrons and therefore trace plasma dynamics on longer timescales, probing very steep electron distributions, whose acceleration mechanisms are still poorly explored \cite{brunetti-jones14}. 
In a longer term perspective, SKA will unlock studies of polarization properties of cluster radio emission, constraining even smaller (resistive?) scales.

{\bf X-rays.}
Over the next decade, significant improvement of X-ray observatories is expected.
The imminent launch of XRISM (early 2022) will open an era of high-resolution X-ray spectroscopy, providing spectra of extended X-ray sources with an energy resolution 30 times better than Chandra’s and XMM-Newton’s. 
Resolving individual spectral lines will allow us to measure ICM turbulence and bulk motions, probe plasma non-equilibrium effects, and elucidate the origin of non-thermal electrons. 
Given the limited spatial resolution of XRISM, the primary focus would be on nearby, bright systems. 
The next generation of X-ray observatories (Athena, Lynx) will also add spatial resolution and effective area to probe the most fundamental properties of weakly-collisional, high-$\beta$ plasmas even in cluster substructures.

{\bf $\gamma$-rays.}
The upcoming Cherenkov Telescope Array (CTA), with its unprecedented sensitivity, will detect or better constrain the cluster TeV emission, both diffuse and from ICM shocks. 

\vspace{-5mm}
\subsubsection{Synergy with Plasma Physics and Broader Impact}
\vspace{-3mm}
Phenomena like collisionless shocks, magnetic turbulence, instabilities driven by energetic particles, and kinetic energy transport are ubiquitous in space and astrophysical plasmas.
Theory, simulations, and observations tackle fundamental processes that span a vast range of scales and environmental parameters, and may be useful also to interpret laboratory plasma experiments.

Connecting with the space/astro communities amplifies the message that plasma physics is crucial for understanding the Universe.
For instance, galaxy formation models have started including magnetic fields and CRs and exoplanet habitability must reckon with stellar winds' plasma properties.
This broad application of plasma concepts increases the need for basic plasma physics training, thereby fostering a new, and more diverse, generation of plasma physicists.

\vspace{-1.7cm}
\setlength{\bibsep}{1pt}
\bibliography{sample}
\bibliographystyle{science}

\end{document}